\documentclass[twoside,11pt]{article}

\usepackage{amssymb}
\usepackage{amsmath}
\usepackage{amsfonts}
\usepackage{wasysym}

\begin{document}           

\title{\large ACCELERATING QUANTUM UNIVERSE}
\author{V.E. Kuzmichev, V.V. Kuzmichev\\[0.5cm]
\itshape Bogolyubov Institute for Theoretical Physics,\\
\itshape National Academy of Sciences of Ukraine, Kiev, 03680 Ukraine}

\date{}

\maketitle

\begin{abstract}
The origin of negative pressure fluid (the dark energy) is investigated
in the quantum model of the homogeneous, isotropic and closed
universe filled with a uniform scalar field and a perfect fluid which
defines a reference frame. The equations of the model are reduced
to the form which allows a direct comparison between them and the equations of 
the Einsteinian classical theory of gravity. It is shown that quantized scalar
field has a form of a condensate which behaves as an antigravitating medium.
The theory predicts
an accelerating expansion of the universe even if the vacuum energy density
vanishes. An antigravitating effect of a condensate has a purely quantum nature.
It is shown that the universe with the parameters close to the Planck ones can
go through the period of exponential expansion.
The conditions under which in semi-classical approximation the universe looks 
effectively like spatially flat with negative deceleration parameter are determined.
The reduction to the standard model of classical cosmology is discussed.
\end{abstract}


\begin{center}
      \textbf{1. Introduction}\\[0.3cm]
\end{center}

An accelerating expansion of the present-day Universe which was discovered
in the late 90th \cite{RP} and confirmed in subsequent type Ia supernova 
observations \cite{TR,WMAP3} gives an evidence that some mysterious component
of the matter-energy (called the dark energy \cite{OSB}) exists in the Universe.
It behaves as an antigravitating medium (fluid). In standard $\Lambda$CDM model
\cite{KolT,OlP,Pee2} this fluid is introduced from the beginning in the form of
the cosmological constant which is identified with the vacuum energy density.
Such a phenomenological approach gives a good description of the observational
data, but it generates the cosmological constant problem \cite{Pee2}.
Nowadays there are several models of the dark energy based on different 
reasonable concepts \cite{CDS,AMS,KMP,Cal,LLK}. It is quite possible that
the origin of the dark energy is connected with quantum processes
in the early Universe.

As is well known quantum theory 
adequately describes properties of various physical systems.
Its universal validity demands that the Universe as a whole 
must obey quantum laws as well. Since  quantum effects are not 
\textit{a priori} restricted to certain scales \cite{Kie}, then one should not 
conclude in advance, without research into the properties of the Universe 
in the theory more general than classical cosmology, 
that they cannot have any impact on processes at large scales 
(the motivation to develop quantum cosmology see \textit{e.g.} in 
\cite{Ish,Cou,KK3}).
One can expect that in semi-classical limit negative pressure fluid
ought to arise as the remnant of the early quantum epoch.

Under construction of quantum theory of gravity one can proceed from the idea that
properties of any quantum system can be described on the basis of the solution of 
some partial differential equation. If a Lagrangian is given then equations 
themselves can be derived from the principle of least action. Passing from the
Lagrangian formalism to the Hamiltonian one it is possible, in principle, to
construct consecutive quantum theory of gravity using the method of canonical
quantization. The first problem on this way is to choose generalized 
variables. Following \textit{e.g.} 
Refs. \cite{And,Whe,DeW}, it is convenient to choose
metric tensor components and matter fields as such variables. But the functional
equations \cite{Whe,DeW} obtained in this approach prove to be insufficiently
suitable for specific problems of quantum theory of gravity and cosmology. These
equations do not contain a time variable in an explicit form. This, in turn, 
gives rise
to the problem of interpretation of the wave function of the universe (see e.g.
discussion in Ref. \cite{KolT} and references therein). A cause of the failure
can be easily understood with the help of Dirac's constraint system theory 
\cite{Dir}. It has been found that the structure of constraints in general relativity 
is such that variables which correspond to true dynamical degrees of freedom cannot
be singled out from canonical variables of geometrodynamics. This difficulty is 
stipulated by an absence of predetermined way to identify spacetime events in 
generally covariant theory \cite{KuchT}.

One way of solving the conceptual problems of theory of gravity
is connected with the introduction of material reference systems
\cite{KuchT,Tor,BroM}. A perfect fluid can determine
a reference frame during consideration of different model systems in classical 
theory \cite{Sch,Lun,Dem,Kij} 
and may appear useful after the passage to the quantum case
as well. Choosing the relativistic matter as the simplest physical realization
of a perfect fluid one can obtain the appropriate equations of quantum 
geometrodynamics for the minisuperspace model \cite{KK2} which allow to study
the properties of the early Universe and to pass to general relativity in
the semi-classical limit.

In this paper the problem of an accelerating expansion of the universe is
analyzed in the framework of the exactly solvable quantum model.
We consider the homogeneous, isotropic and closed universe filled with a uniform
scalar field and the relativistic matter which defines a reference frame.

In Section 2 the basic concepts and the equations of the quantum 
model of the universe obtained in Ref. \cite{KK2} are introduced.
In Section 3 the canonical equations which determine the time change of the scale 
factor, scalar field and the operators of the momenta canonically conjugate with
these two variables are obtained.
In Section 4 the quantum analogues of the Einstein-Friedmann equations are given
and the conclusion is drawn about an antigravitating effect of
a condensate of excitation quanta of oscillations of a primordial 
scalar field about an equilibrium state.
In Section 5 the semi-classical limit is considered and the Einstein-Friedmann 
equations are obtained. 
In Section 6 the properties of the universe with large masses of a condensate
in the state of the true vacuum of scalar field are studied. 
The conditions under which the universe can undergo the phase of exponential
expansion (inflation) and look effectively like spatially flat with negative deceleration parameter are established.
Section 7 presents some concluding remarks. 

In this paper we use the modified Planck system of units in which 
$l_{P} = \sqrt{\frac{2 G \hbar}{3 \pi c^{3}}}$ is taken as a unit of length,
$\rho_{P} = \frac{3 c^{4}}{8 \pi G l_{P}^{2}}$ is a unit of energy density and so on.
All relations are written for dimensionless values.

\begin{center}
      \textbf{2. Equations of quantum model of the universe}\\[0.3cm]
\end{center}

Let us consider the homogeneous, isotropic and closed universe which is described
by the Robertson-Walker metric
\begin{equation}\label{01}
    ds^{2} = a^{2}(\eta)\,[N^{2}(\eta)d\eta^{2} - d\Omega^{2}],
\end{equation}
where $a$ is the cosmic scale factor, $N$ is the lapse function that specifies
the time reference scale, $d\Omega^{2}$ is an interval element on a unit 
three-sphere, $\eta$ is the time variable (conformal time at $N = 1$). Let us suppose
that the universe is filled with the uniform scalar field $\phi$ with the potential
energy density (potential) $V(\phi)$ and a perfect fluid which defines material 
reference frame. The Hamiltonian $H$ of such a system has the form of a linear 
combination of constraints \cite{KK2} and weakly vanishes (in Dirac's sence 
\cite{Dir}),
\begin{eqnarray}\label{02}
    H = \frac{N}{2}\left\{-\,\pi_{a}^{2} - a^{2} + \frac{2}{a^{2}}\,\pi_{\phi}^{2} + a^{4} [V(\phi) + \rho]\right\} \nonumber \\
+ \lambda_{1}\left\{\pi_{\Theta} - \frac{1}{2}\,a^{3} \rho_{0} s\right\} 
+ \lambda_{2}\left\{\pi_{\tilde{\lambda}} + \frac{1}{2}\,a^{3} \rho_{0} \right\} \approx 0,
\end{eqnarray}
where $\rho = \rho(\rho_{0}, s)$ is the energy density of a perfect fluid, $\rho_{0}$ 
is the density of the rest mass, $s$ is the specific entropy. 
The $\Theta$ is the thermasy 
(potential for the temperature, $T = \Theta_{,\,\nu} U^{\nu}$). 
The $\tilde{\lambda}$ is the potential for the specific Gibbs free energy $f$ 
taken with an inverse sign, 
$f = -\,\tilde{\lambda}_{,\,\nu} U^{\nu}$. 
The $U^{\nu}$ is the four-velocity. 
The $\pi_{a},\, 
\pi_{\phi},\, \pi_{\Theta},\, \pi_{\tilde{\lambda}}$ are the momenta canonically
conjugate with the variables $a,\, \phi,\, \Theta,\, \tilde{\lambda}$ respectively. 
The momenta $\pi_{\rho_{0}}$ and $\pi_{s}$
conjugate with the variables $\rho_{0}$ and $s$ vanish identically. The 
$N$, $\lambda_{1}$, and $\lambda_{2}$ are Lagrange multipliers.

From the conservation of primary constraints in time it follows
the conservation laws
\begin{equation}\label{3}
    E_{0} \equiv a^{3} \rho_{0} = \mbox{const}, \qquad s =  \mbox{const},
\end{equation}
where the first relation describes the conservation law of a
macroscopic value which characterizes the number of particles. For example,
if a perfect fluid is composed of baryons, then this condition reflects
the conservation of baryon number. The second equation in (\ref{3}) 
represents the conservation of the specific entropy.

Taking into account these conservation laws 
and vanishing of the momenta $\pi_{\rho_{0}}$ 
and $\pi_{s}$, one can discard degrees of freedom corresponding to the variables
$\rho_{0}$ and $s$, and convert the second-class constraints into
first-class constraints in accordance with Dirac's proposal \cite{Dir}.
We have used the same approach in Ref. \cite{KK2}. In quantum theory 
first-class constraint equations become constraints on the wave function $\Psi$. 

It is convenient to pass from the generalized variables $\Theta$ and 
$\widetilde{\lambda}$ to the non-coordinate co-frame
\begin{eqnarray}\label{03}
  h\, d\tau = s\,d\Theta\, -\,d\widetilde{\lambda}, \nonumber \\
  h\, dy = s\,d\Theta\, +\,d\widetilde{\lambda},
\end{eqnarray}
where $h = \frac{\rho + p}{\rho_{0}}$ is the specific enthalpy which plays the role
of inertial mass \cite{Sch}, $p$ is the pressure, $\tau$ is proper time in every 
point of space. The corresponding derivatives commute 
between themselves,
$$
\left[\partial_{\tau},\,\partial_{y}\right] = 0.
$$
Thus the quantum universe in which a perfect fluid is taken in the form of 
relativistic matter is described by the equations \cite{KK2}
\begin{equation}\label{1}
    \left\{-\, i\,\partial_{\tau_{c}} - \frac{1}{2}\, E_{0} \right\} \Psi = 0,
\end{equation}
\begin{equation}\label{2}
    \left\{-\,\partial^{2}_{a} + \frac{2}{a^{2}}\, \partial^{2}_{\phi} +  a^{2} - 
a^{4} V(\phi) - E\right\} \Psi = 0,
\end{equation}
where $\Psi$ does not depend on the variable $y$, 
$E \equiv a^{4} \rho = \mbox{const}$,
$\rho$ is the energy density of relativistic matter. 
The value $\tau_{c}$ is the time variable connected
with the proper time $\tau$ by the differential relation,  $d\tau_{c} = h\, d\tau$.
Eq. (\ref{1}) has a particular solution in the form
\begin{equation}\label{5}
    \Psi = \mbox{e}^{i E \bar{\tau}} |\psi \rangle,
\end{equation}
where we have changed the time scale using the relation 
$E \bar{\tau} = \frac{1}{2} E_{0} \tau_{c}$.
The vector $|\psi\rangle$ is defined in the space of two variables  $a$ and $\phi$,
and determined by Eq. (\ref{2}) which we rewrite in the form
\begin{equation}\label{6}
    \left(\hat{\pi}_{a}^{2} + a^{2}- 2\, a\,\hat{H}_{\phi} 
- E\right)|\psi\rangle = 0,
\end{equation}
where $\hat{\pi}_{a} = -\,i\,\partial_{a}$ is the operator of the 
momentum canonically conjugate with the variable $a$,
\begin{equation}\label{7}
    \hat{H}_{\phi} = \frac{1}{2}\,a^{3}\hat{\rho}_{\phi}
\end{equation}
is the operator of mass-energy of a scalar field in a comoving volume 
$\frac{1}{2}\, a^{3}$.
Here
\begin{equation}\label{8}
    \hat{\rho}_{\phi} = \frac{2}{a^{6}}\,\hat{\pi}_{\phi}^{2} + V(\phi)
\end{equation}
is the operator of the energy density of a scalar field, 
$\hat{\pi}_{\phi} = -\,i\,\partial_{\phi}$ is the operator of the momentum
canonically conjugate with the variable $\phi$.

Eqs. (\ref{5}) and (\ref{6}) are equivalent\footnote{These equations
can be obtained even without an introduction of proper time $\tau$
by building a time variable from the matter variables (cf. \cite{Lun,Dem,Kij}), 
\textit{e.g.} considering the thermasy $\Theta$ as a time variable.} 
to the Schr\"{o}dinger-type
equation which was obtained in Refs. \cite{Kuz1,KK1}
within the bounds of the scheme \cite{KuchT} for incorporating of a reference
systems in general relativity through the imposition of coordinate
conditions before variation of the action.

Eq. (\ref{6}) can be integrated with respect to $\phi$, if one determines the
form of the potential $V(\phi)$. 
As in Ref. \cite{KK2} we consider the solution of Eq.
(\ref{6}) when the field $\phi$ is near its minimum at the point 
$\phi = \sigma$. Then $V(\phi)$ can be approximated by the expression 
\begin{equation}\label{11}
    V(\phi) = \rho_{\sigma} + \frac{{m}_{\sigma}^{2}}{2}\, (\phi - \sigma)^{2},
\end{equation}
where $\rho_{\sigma} = V(\sigma)$, ${m}_{\sigma}^{2} = 
[d^{2}V(\phi)/d\phi^{2}]_{\sigma} > 0$.
If $\phi = \sigma$ is the point of absolute minimum, then $\rho_{\sigma} = 0$ and
the state $\sigma$ corresponds to the true vacuum of a primordial scalar field, 
while the state with $\rho_{\sigma} \neq 0$ matches with the false vacuum \cite{Col}. 

Introducing the new variable
\begin{equation}\label{12}
    x = \left(\frac{m_{\sigma} a^{3}}2{}\right)^{1/2} (\phi - \sigma),
\end{equation}
which describes a deviation of the field $\phi$ from its equilibrium state,
we find that 
\begin{equation}\label{13}
    \hat{H}_{\phi} |u_{k}\rangle = \left(M_{k} + \frac{1}{2}\,a^{3} 
\rho_{\sigma}\right)|u_{k}\rangle,
\end{equation}
where $\langle x |u_{k}\rangle$ are the functions of harmonic oscillator 
with $k = 0,\,1,\,2,\,... $, and
\begin{equation}\label{14}
     M_{k} = m_{\sigma} \left(k + \frac{1}{2}\right).
\end{equation}
The quantity $M_{k}$ describes an amount of matter (mass) in the 
universe related to a scalar field.
This mass is represented in the form of a sum of masses of excitation quanta of 
the spatially coherent oscillations of the field $\phi$ about the equilibrium 
state $\sigma$, $k$ is the number of quanta. The mentioned
oscillations correspond to a condensate of zero-momentum $\phi$ quanta with the
mass $m_{\sigma}$. 
If $\rho_{\sigma} \neq 0$, then the value $\frac{1}{2}\,a^{3} \rho_{\sigma}$
is the energy of the false vacuum in the universe with the scale factor $a$.

We shall look for the solution of Eq. (\ref{6}) in the form of the superposition of
the states with different masses $M_{k}$ 
\begin{equation}\label{15}
    |\psi\rangle = \sum_{k}\, |f_{k}\rangle |u_{k}\rangle.
\end{equation}
Using orthonormality of the states  $|u_{k}\rangle$ we obtain the equation for
the vector $|f_{k}\rangle$
\begin{equation}\label{16}
     \left(\hat{\pi}_{a}^{2} + a^{2} - 2 a M_{k} - a^{4} \rho_{\sigma} - E\right) |f_{k}\rangle = 0.
\end{equation}
In the case $\rho_{\sigma} = 0$ this equation is exactly integrable \cite{KK2}. 
At the same time the state $\Psi$ has a finite norm, $\langle \Psi|\Psi\rangle < \infty$ (cf. \textit{e.g.} Ref. \cite{KuchT}).

\begin{center}
    \textbf{3. Canonical equations}\\[0.3cm]
\end{center}

The equation of motion for an arbitrary function $\mathcal{O}$ 
of the variables $a,\,\phi,\, 
\pi_{a}$ and $\pi_{\phi}$ has the form
\begin{equation}\label{4}
    \frac{d\mathcal{O}}{d\eta} \approx \{\mathcal{O}, H\},
\end{equation}
where $H$ is the Hamiltonian (\ref{02}), the sign $\approx$ means that
Poisson brackets $\{.,.\}$ must all be worked out 
before the use of the constraint equations. 
In quantum theory the equation of motion for an operator $\hat{\mathcal{O}}$ 
in the Heisenberg representation takes the form
\begin{equation}\label{18}
     \frac{d\hat{\mathcal{O}}}{d\eta} \approx \frac{1}{i}\,
[\hat{\mathcal{O}},\hat{H} ],
\end{equation}
where $[.,.]$ is a commutator, and $\hat{H}$ is determined by the expression
(\ref{02}), in which all dynamical variables are substituted with operators.

Let $\hat{\mathcal{O}} = a$, then from Eq. (\ref{18}) we obtain 
\begin{equation}\label{21}
      \frac{d\hat{a}}{d\eta} = -\,N\,\hat{\pi}_{a}.
\end{equation}
This relation can be considered as a definition of 
the momentum operator $\hat{\pi}_{a}$.
Let $\hat{\mathcal{O}} = \hat{\pi}_{a}$. Then from Eq. (\ref{18}) it follows that
\begin{equation}\label{23}
    \frac{1}{N}\,\frac{d\hat{\pi}_{a}}{d\eta} = 
\frac{2}{a^{3}}\,\hat{\pi}_{\phi}^{2}\, +\,a\, 
-\,2\,a^{3} V(\phi).
\end{equation}
If $\hat{\mathcal{O}} = \phi$, then
\begin{equation}\label{24}
    \frac{1}{N}\,\frac{d\phi}{d\eta} = \frac{2}{a^{2}}\,\hat{\pi}_{\phi}.
\end{equation}
If $\hat{\mathcal{O}} = \hat{\pi}_{\phi}$, then
\begin{equation}\label{024}
    \frac{1}{N}\,\frac{d\hat{\pi}_{\phi}}{d\eta} = -\,
 \frac{a^{4}}{2}\,\frac{dV(\phi)}{d\phi}.
\end{equation}

\begin{center}
     \textbf{4. Quantum analogues of the Einstein-Friedmann equations}
\\[0.3cm]
\end{center}

Eq. (\ref{16}) can be written in the form
\begin{equation}\label{25}
    \left\{\left(\frac{1}{a^{2}}\,\hat{\pi}_{a}\right)^{2}\, -\,\rho_{tot}\, 
+\,\frac{1}{a^{2}}\right\} |f_{k}\rangle = 0,
\end{equation}
where the quantity
\begin{equation}\label{26}
    \rho_{tot} = \rho_{k}\, +\,\rho\, +\,\rho_{\sigma}
\end{equation}
with the components
\begin{equation}\label{27}
    \rho_{k} \equiv \frac{2 M_{k}}{a^{3}}, \qquad \rho \equiv \frac{E}{a^{4}}
\end{equation}
can be interpreted as a total energy density. 
It represents itself the sum of the energy densities of a condensate
of $\phi$ quanta $\rho_{k}$, relativistic matter $\rho$ and the false vacuum 
$\rho_{\sigma}$.

In order to obtain the second equation of quantum model we use the 
operator equation (\ref{23}) rewritten in the equivalent form
\begin{equation}\label{28}
    \frac{1}{a^{3}N}\,\frac{d\hat{\pi}_{a}}{d\eta} - \hat{\rho}_{\phi}\, 
+\,3\,V(\phi)\, -\,\frac{1}{a^{2}} = 0.
\end{equation}
Acting with this operator equation 
on $|\psi\rangle$, taking into account Eqs. (\ref{7}), (\ref{11}), (\ref{13}), 
(\ref{15}) and
\begin{eqnarray}\label{29}
   V(\phi) |u_{k}\rangle = \left(\frac{1}{a^{3}}\,M_{k}\, 
+\,\rho_{\sigma}\right)|u_{k}\rangle\, +\,\frac{1}{2 a^{3}}\sqrt{\left(M_{k}\, 
-\,\frac{m_{\sigma}}{2}\right)\left(M_{k}\, -\,\frac{3 m_{\sigma}}{2}\right)}\,\, 
|u_{k-2}\rangle \nonumber \\
+\,\frac{1}{2 a^{3}}\sqrt{\left(M_{k}\, +\,\frac{m_{\sigma}}{2}\right)\left(M_{k}\,
 +\,\frac{3 m_{\sigma}}{2}\right)}\,\, |u_{k+2}\rangle,
\end{eqnarray}
we obtain
\begin{equation}\label{30}
   \left\{\frac{1}{a^{3}N}\,\frac{d\hat{\pi}_{a}}{d\eta}\,
+\,\frac{1}{2}\,\rho_{k}\, 
+\,2\,\rho_{\sigma}\, -\,\frac{1}{a^{2}}\right\}|f_{k}\rangle + 
\sum_{k'}\,\mathcal{P}_{kk'} |f_{k'}\rangle = 0, 
\end{equation}
where the operator
\begin{eqnarray}\label{32}
    \mathcal{P}_{kk'} = \frac{3}{2 a^{3}}\left\{\sqrt{\left(M_{k}\, 
+\,\frac{m_{\sigma}}{2}\right)\left(M_{k}\, +\,\frac{3 m_{\sigma}}{2}\right)}\,\, 
\delta_{k',k+2}\,\right.\nonumber  \\
\left.+\,\sqrt{\left(M_{k}\, -\,\frac{m_{\sigma}}{2}\right)\left(M_{k}\, -\,\frac{3 
m_{\sigma}}{2}\right)}\,\, \delta_{k',k-2}\right\}
\end{eqnarray}
takes into account the quantum effects stipulated by 
the character of the wavefunction (\ref{15}) which is the superposition of 
$|f_{k}\rangle$ corresponding to different masses of a condensate. 

Eqs. (\ref{25}) and (\ref{30}) are the quantum analogues of the Einstein-Friedmann 
equations in general relativity. In the approximation 
$\mathcal{P}_{kk'} = 0$ Eq. (\ref{30}) can be represented in the ``standard''
form for the perfect fluid source
\begin{equation}\label{33}
    \left\{\frac{1}{a^{3}N}\,\frac{d\hat{\pi_{a}}}{d\eta}\,
+\,\frac{1}{2}\left(\rho_{tot}\, -\,3\,p'_{tot}\right) - 
\frac{1}{a^{2}}\right\}|f_{k}\rangle = 0,
\end{equation} 
where $\rho_{tot}$ is the density (\ref{26}), while
\begin{equation}\label{34}
   p'_{tot} = p'_{k}\, +\,p\, +\,p_{\sigma}
\end{equation}
with the components
\begin{equation}\label{35}
   p'_{k} = 0, \qquad p = \frac{1}{3}\,\rho \qquad p_{\sigma} = -\,\rho_{\sigma}
\end{equation}
is the isotropic pressure (here a dash marks the fact that the 
sum over $k'$ in Eq. (\ref{30}) is not taken into account). 
From Eq. (\ref{35}) one can conclude that 
the constant component $\rho_{\sigma}$
of the energy density (\ref{26}) is described by the vacuum-type equation of 
state, while a condensate of $\phi$ quanta
in the case under consideration
is a perfect fluid with zero pressure. But taking into account 
the presence of operator $\mathcal{P}_{kk'}$ in Eq. (\ref{30})
leads to the considerable modification of physical properties of
a condensate.

Let us note the result (\ref{35}) is non-trivial. Starting from a uniform
scalar field, after quantization we obtain the pressureless matter component 
(dust).

We shall study the role of $\mathcal{P}_{kk'}\neq 0$ in Eq. (\ref{30}). 
In the limit of large values of $k$ we have 
$M_{k} \gg \frac{3}{2}\,m_{\sigma}$,
and the operator $\mathcal{P}_{kk'}$ takes the form
\begin{equation}\label{36}
    \mathcal{P}_{kk'} = \frac{3}{2}\, \rho_{k}\,\delta_{kk'} \qquad \mbox{at} \qquad 
k \gg 1.
\end{equation}
Substituting Eq. (\ref{36}) into (\ref{30}) we obtain the equation 
\begin{equation}\label{37}
\left\{\frac{1}{a^{3}N}\,\frac{d\hat{\pi_{a}}}{d\eta}\,+\,\frac{1}{2}
\left(\rho_{tot}\, -\,3\,p_{tot}\right)- \frac{1}{a^2} \right\}|f_{k}\rangle = 0,
\end{equation}
where the pressure is equal to
\begin{equation}\label{39}
     p_{tot} = p_{k}\, +\,p\, +\,p_{\sigma},
\end{equation}
and the equations of state for relativistic matter and vacuum are the same as in Eq. 
(\ref{35}), but the equation of state of a condensate takes the form
\begin{equation}\label{40}
    p_{k} = -\, \rho_{k}.
\end{equation}
This means that taking into account the quantum effects caused by the non-zero
operator $\mathcal{P}_{kk'}$
at large values of masses $M_{k}$ leads to the
fact that a condensate of $\phi$ quanta obtains the property of an antigravitating
medium. As a result the quantum universe will expand in an accelerating mode at
$\rho_{k} > \rho$ even in the state of true vacuum of the field $\phi$ 
($\rho_{\sigma} = 0$).

\begin{center}
    \textbf{5. Semi-classical limit}
\\[0.3cm]
\end{center}

We shall take the wave function $\langle a|f_{k} \rangle$ in the form
\begin{equation}\label{040}
  \langle a|f_{k} \rangle = A(a)\,\mbox{e}^{i S(a)},
\end{equation}
where $A$ and $S$ are some real functions of $a$ and the index $k$ is omitted for
simplicity. Substituting Eq. (\ref{040}) into (\ref{25}) we obtain the equivalent
equation
\begin{equation}\label{041}
   \frac{1}{a^{4}} \left(\partial_{a} S \right)^{2} - \rho_{tot} - 
\frac{1}{a^{4}}\,\frac{\partial_{a}^{2} A}{A} + \frac{1}{a^{2}} -
\frac{i}{a^{4} A^{2}}\,\partial_{a} \left(A^{2} \partial_{a} S \right) = 0.
\end{equation}
The imaginary part gives the continuity equation. Eq. (\ref{041}) written in the
ordinary units takes the form
\begin{eqnarray}\label{042}
   \frac{1}{a^{4}} \left(\partial_{a} S \right )^{2} - 
\frac{8\pi G}{3c^{4}}\,\rho_{tot} &-& 
\left(\frac{2G \hbar}{3\pi c^{3}} \right)^{2}\,
\frac{1}{a^{4}}\,\frac{\partial_{a}^{2} A}{A} + \frac{1}{a^{2}} \nonumber \\
 &-& i\,\frac{2G \hbar}{3\pi c^{3}}\,
\frac{1}{a^{4} A^{2}}\,\partial_{a} \left (A^{2} \partial_{a} S \right ) = 0.
\end{eqnarray}
Neglecting the terms which are proportional to $\hbar$ and $\hbar^{2}$ we get
the classical Hamilton-Jacobi equation for the action $S$. The classical momentum 
$\pi_{a}$ is equal to
\begin{equation}\label{043}
   \pi_{a} = \partial_{a} S = -a\,\frac{da}{d\tau},
\end{equation}
where $d\tau = a N d\eta$, $\tau$ is proper time.

Substituting Eq. (\ref{040}) into (\ref{37}) and taking into account the rule for the 
derivative of an operator $\mathcal{O}$,
$$
\langle f_{k} | \frac{1}{N}\,\frac{d \mathcal{O}}{d \eta} | 
f_{k} \rangle =  
\frac{d}{N d \eta}\,\langle f_{k} | \mathcal{O} | f_{k} \rangle,
$$
we obtain
\begin{equation}\label{044}
   \frac{1}{a^{3} N}\,\frac{d}{d \eta}\,\left(\partial_{a} S \right) + 
\frac{1}{2}\,\left(\rho_{tot} - 3 p_{tot} \right) - \frac{1}{a^{2}} -
\frac{i}{a^{3} N}\,\frac{d}{d \eta}\,\left(\frac{\partial_{a} A}{A} \right) = 0.
\end{equation}
The imaginary part of this equation taken in the ordinary units is proportional to 
$\hbar$.

Omitting terms proportional to $\hbar$ and $\hbar^{2}$ in Eqs. (\ref{041}) and 
(\ref{044}) and taking into account Eq. (\ref{043}), we obtain the Einstein-Friedmann
equations
\begin{equation}\label{77}
    \left(\frac{1}{a}\,\frac{d a}{d\tau}\right)^{2} = \rho_{tot}\, 
-\,\frac{1}{a^{2}}, \qquad 
\frac{1}{a}\,\frac{d^{2} a}{d\tau^{2}} = 
-\,\frac{1}{2}\,\left(\rho_{tot}\, +\,3\, p_{tot}\right),
\end{equation}
where the total energy density $\rho_{tot}$ and the pressure $p_{tot}$ are 
determined 
in Eqs. (\ref{26}) and (\ref{39}) with the equation of state of a condensate
(\ref{40}). It means that the conclusion about 
an antigravitating effect of a condensate made above 
remains valid after a passage to the classical limit as well.
Let us note that the second equation is obtained from Eq. (\ref{37}) and it
contains the contribution from the operator  $\mathcal{P}_{k k'}$ (\ref{36}).
If the quantum effects 
determined by this operator had been omitted, the equation of state of a condensate
would have the form $p_{k} = 0$ as for a dust.

From the equations (\ref{77}) it follows the equation for the evolution of 
$\rho_{tot}$,
\begin{equation}\label{770}
    \frac{d \rho_{tot}}{d\tau} = 
           - 4 \left(\frac{1}{a}\,\frac{d a}{d\tau}\right) \rho.
\end{equation} 

In classical theory the value $E = a^{4} \rho$ is a positive constant which is
determined by the values of $a$ and $\rho$ at a certain instant of time. In quantum
theory it is quantized in accordance with Eq. (\ref{16}). We shall study the
effect of quantization of $E$ on the properties of the universe.

\begin{center}
    \textbf{6. Universe in the state of true vacuum of 
scalar field}\\[0.3cm]
\end{center}

Let us consider the exactly solvable model with $\rho_{\sigma} = 0$ and 
large masses of a condensate, $M_{k} \gg 1$. Eq. (\ref{25}) is strictly 
equivalent to the stationary equation (\ref{16}). 
The state vector  $|f_{k}\rangle$ is characterized 
by the additional index $n$ which numbers the states of the universe in the 
effective potential well
\begin{equation}\label{440}
    U(a) = a^{2} - 2 a M_{k}.
\end{equation}

Solving Eq. (\ref{16}) for the case under consideration 
with the boundary conditions $\langle 0|f_{n,k}\rangle \neq 0$\footnote{The choice
of boundary conditions at the point $a = 0$ should be stipulated by the physical
properties of the system under study. In classical cosmology
the universe near the initial cosmological singularity point
$a = 0$ is characterized by nontrivial values of energy density. Since
$\left|\langle 0|f_{n,k}\rangle\right|^{2}$ is the particle number density at $a = 0$,
then the choice of such a boundary condition is justified from the
cosmological point of view.} and $\langle a|f_{n,k}\rangle \rightarrow 0$ at 
$a \rightarrow \infty$ 
we find the state vectors in $a$-representation \cite{KK2}
\begin{equation}\label{42}
    \langle a|f_{n,k}\rangle = N_{n,k}\,\mbox{e}^{- 
\frac{1}{2}\,(a - M_{k})^{2}}\,H_{n}(a - M_{k}),
\end{equation}
where $H_{n}(\xi)$ are Hermitian polynomials, $n = 0,1,2, \ldots$, and
\begin{eqnarray}\label{43}
    N_{n,k} &=& \left\{2^{n-1}n!\sqrt{\pi}\left(\mbox{erf} M_{k} + 1\right) 
\right. \nonumber \\
 & & \left. - \mbox{e}^{- M_{k}^{2}}\, \sum _{l=0}^{n-1}\, \frac{2^{l}n!}{(n-l)!}\, 
H_{n-l}(M_{k})\,H_{n-l-1}(M_{k})\right\}^{- \frac{1}{2}}
\end{eqnarray}
is the normalization factor,  $\mbox{erf} M_{k}$ is the probability integral.  
These state vectors correspond to the eigenvalues
\begin{equation}\label{41}
    E = 2n + 1 - M_{k}^{2}.
\end{equation}
The quantum-mechanical mean value 
\begin{equation}\label{44}
    \langle a \rangle_{n,k} = \langle f_{n,k}|a|f_{n,k}\rangle
\end{equation}
in the state (\ref{42}) is equal to
\begin{equation}\label{45}
    \langle a \rangle_{n,k} = M_{k} + \langle \xi \rangle_{n,k},
\end{equation}
where
\begin{eqnarray}\label{46}
    \langle \xi \rangle_{n,k} & = & 
 \int_{-M_{k}}^{\infty}\! d\xi \,\xi\, \langle \xi + 
M_{k}|f_{n,k}\rangle^{2} \nonumber \\
& = & N_{n,k}^{2}\, 2^{n-1} n!\, \mbox{e}^{- M_{k}^{2}} \left\{1 + 
\sum_{l=0}^{n-1}\, \frac{2^{l-n}}{(n-l)!}\,H_{n-l}(M_{k})\,H_{n-l-1}(M_{k}) \right\}
\end{eqnarray}
is the correction which can be neglected in the case $M_{k} \gg 1$.

In Ref. \cite{KK2} it was shown that the quantum universe can nucleate from
the initial cosmological singularity point $a = 0$ with the non-zero probability.
The universe being nucleated in the ground ($n = 0$) state has the Planck parameters,
$M_{k} = 1$ and $\langle a \rangle_{0,k} = 1.11$. In this state $E = 0$ and 
according to Eq. (\ref{770}) $\rho_{tot} = \mbox{const}$. After integrating
the second equation in (\ref{77}) we find that the universe in such a state will
expand exponentially
\begin{equation}\label{771}
    a(\tau) = a(0) \exp \left\{\sqrt{\rho_{tot}} \tau\right\}.
\end{equation}
This exponential expansion lasts as long as the condition $E = 0$ is satisfied. In
the models of inflation this time is about $10^{-32} - 10^{-35}$ sec \cite{KolT}.

Let the condition $E = 0$ be realized at some instant $\tau = \tau_{2}$ of the evolution of the early universe. The mass of a condensate and the scale factor here take the values $M_{k_{2}}$ and $a_{2}$ respectively, and the total energy density
is equal to 

$$\rho_{tot} = \rho_{k_{2}} = \frac{2M_{k_{2}}}{a_{2}^{3}}.$$

Introducing the Hubble expansion rate $H$, the deceleration parameter $q$ and
the density parameters $\Omega_{tot}$ and $\Omega_{k}$ 
in chosen system of units at some instant $\tau = \tau_{i}$ we have 
\begin{equation}\label{88}
    H_{i} = \left(\frac{1}{a}\,\frac{d a}{d\tau}\right)_{i}, \quad 
q_{i} = - \frac{1}{H^{2}_{i}} \left(\frac{1}{a}\frac{d^{2} a}{d\tau^{2}}\right)_{i}, \quad \Omega_{tot} = \frac{\rho_{tot}}{H^{2}_{i}}, \quad
\Omega_{k_{i}} = \frac{\rho_{k_{i}}}{H^{2}_{i}}.
\end{equation}

Then from (\ref{77}) we obtain
\begin{equation}\label{512}
    \Omega_{tot} = 1, \qquad q_{2} = - 1,
\end{equation}
where we take into account that under consideration of the early universe the curvature 
term $a^{-2}_{2}$ can be dropped. Hence the early universe is accelerating.

In the quantum model under study the evolution of the universe is described 
as transitions with the non-zero probabilities between the states of the universe with
different masses of a condensate \cite{KK2}. An increase in this mass
leads to an expansion of the universe. Breaking of the condition
$E = 0$ results in change from exponential expansion to power law.

Let us consider the case $E \neq 0$ and $M_{k} \gg 1$. We choose the 
instant of time $\tau_{1}$ for which the scale factor $a(\tau_{1}) \equiv a_{1}$
satisfies the condition $a_{1} = M_{k_{1}}$ \footnote{The equations (\ref{77}) are 
equivalent to the Hamilton equations of classical cosmology. They do not
take into account the dispersions of $a$ and the conjugate momentum $\pi_{a}$.
According to the Ehrenfest theorem one can replace in (\ref{77}) the
classical value of the scale factor $a$ with the mean value (\ref{45})
at every instant of time.}. 
Then taking into account (\ref{41}) from (\ref{77}) we obtain
\begin{equation}\label{84}
    \left(\frac{1}{a} \frac{da}{d\tau}\right)^{2}_{1} = \rho_{n_{1}},  
\qquad 
\left(\frac{1}{a} \frac{d^{2} a}{d\tau^{2}}\right)_{1} = \frac{3}{2} \rho_{k_{1}} 
- \rho_{n_{1}},
\end{equation}
where 
\begin{equation}\label{82}
   \rho_{n_{1}} = \frac{2 n + 1}{a^{4}_{1}}, \qquad \rho_{k_{1}} = \frac{2}{a^{2}_{1}}.
\end{equation}
These relations can be interpreted as the Einstein-Friedmann equations 
(at fixed instant $\tau_{1}$) for the
spatially flat universe with the total energy density 
$\rho_{tot} = \rho_{n_{1}}$ and the pressure 
$p_{tot} = p_{k_{1}} + p_{n_{1}}$, where 
$p_{k_{1}} = - \rho_{k_{1}}$, $p_{n_{1}} = \frac{1}{3}\,\rho_{n_{1}}$ 
\footnote{In this case the term with curvature $- a^{-2}$ in (\ref{77}) cannot
be distinguished from the term $\rho_{k}$. Effectively the curvature makes
a kind of renormalization of density of a condensate whose action is totally
compensated in the first relation (\ref{84}). But the second relation contains
a trail from a condensate with vacuum-type equation of state.}.

If we set $\mathcal{P}_{kk'} = 0$, the second equation reduces to
\begin{equation}\label{85}
    \left(\frac{1}{a}\,\frac{d^{2} a}{d\tau^{2}}\right)_{1} = - \rho_{n_{1}}.
\end{equation}
It implies that in the classical limit with $\rho_{\sigma} = 0$, which does not take
into account the quantum effects caused by $\mathcal{P}_{kk'}$, an expansion of the 
universe will be decelerating at any instant of time for which $a_{1} = M_{k_{1}}$, 
while quantum model predicts a possibility
of an accelerating expansion at $\frac{3}{2}\,\rho_{k_{1}} > \rho_{n_{1}}$ even
when $\rho_{\sigma} = 0$.

The relations (\ref{84}) can be rewritten as
\begin{equation}\label{89}
   \Omega_{tot} = 1,\qquad q_{1} = 1 - \frac{3}{2}\,\Omega_{k_{1}}.
\end{equation}
In the epoch when $\Omega_{k_{1}} > \frac{2}{3}$
the expansion of the universe will be accelerating due to antigravitating effect
of a condensate.

According to Eqs. (\ref{84}) and (\ref{82}) the instant $\tau = \tau_{1}$ 
can be associated with the radiation-dominated epoch. In order to consider
the matter-dominated era the additional conjectures about the production of ordinary 
matter from a condensate are required.

\begin{center}
    \textbf{7. Concluding remarks}\\[0.3cm]
\end{center}

This quantum model of the universe can be reduced to the standard model of classical
cosmology in the limit of large quantum numbers.
We suppose \cite{KK5} that ordinary matter is produced in the decay of $\phi$ 
quanta to dark matter particles, baryons, leptons or to their antiparticles. 
Particles and antiparticles can annihilate between themselves and contribute to 
the observed
cosmic microwave background radiation with excess of $\gamma$ quanta over matter
(at a ratio $\eta = n_{B}/n_{\gamma} \sim 10^{-10}$ in our Universe). The part of
a condensate which does not decay to the instant of observation $\tau_{0}$ 
forms dark energy. Using the energy conservation law, one can write 
\begin{equation}\label{90}
    M_{k_{0}} = M_{m} + M_{\Lambda} + M_{\gamma} + Q,
\end{equation}
where $M_{m}$ is the total mass of all baryons, leptons and dark matter, 
$M_{\Lambda} = \frac{1}{2}\,a^{3}_{0} \rho_{\Lambda}$ 
is the mass of a condensate which does not decay
in a comoving volume $\frac{1}{2}\,a^{3}_{0}$ 
with the energy density $\rho_{\Lambda}$ 
and the equation of state $p_{\Lambda} = - \rho_{\Lambda}$, 
$M_{\gamma} = \frac{1}{2} E_{\gamma} a^{-1}_{0}$ 
is the mass of relativistic matter produced by decaying condensate,
$E_{\gamma} = \mbox{const}$, $Q$ is the 
total kinetic energy of relative motion of decay products of $\phi$ quanta. From
Eq. (\ref{90}) it follows the representation for the density $\rho_{k_{0}}$
\begin{equation}\label{91}
    \rho_{k_{0}} = \frac{2 M_{k_{0}}}{a^{3}_{0}} =
2 \frac{M_{m} + Q}{a^{3}_{0}} + \rho_{\Lambda} 
+ \frac{E_{\gamma}}{a^{4}_{0}}.
\end{equation}
In the approximation $Q \sim 0$ (when this term is small in comparison with other 
summands in Eq. (\ref{90})) we obtain the standard cosmological model 
\cite{KolT,OlP,Pee2}. The numerical estimations of $\Omega_{m}$ and 
$\Omega_{\Lambda}$ made in the model (\ref{90}) are 
in agreement with observational data \cite{KK5}.

In order to estimate the value of the decelerating parameter $q_{0}$ in the 
matter-dominated epoch we choose the instant of time $\tau = \tau_{0}$
for which the universe is in the state with $E = M_{k_{0}}^{2}$. This state
is characterized by the quantum number $n_{0} \simeq M_{k_{0}}^{2} \gg 1$
and the condition $a_{0} = M_{k_{0}}$. Then from (\ref{77}) we obtain
\begin{equation}\label{92}
    \Omega_{tot} = 1, \qquad q_{0} = - \frac{1}{2},
\end{equation}
where $\rho_{tot} = \rho_{k_{0}}$.
These equations demonstrate that in the state under consideration 
the curvature term  $a^{-2}$ compensates for the density $\rho$ and 
the universe looks like spatially flat.

The WMAP3 data \cite{WMAP3} give $\Omega_{tot} = 1.003_{-0.017}^{+0.013}$ and
$q_{0} = - 0.63_{-0.07}^{+0.07}$ for the present-day Universe. These values
correspond to the matter energy density $\Omega_{m} = 0.24_{-0.04}^{+0.03}$
and the dark energy density $\Omega_{\Lambda} = 0.76_{-0.06}^{+0.04}$.

If we suppose that the properties of
our Universe are described by the quantum model considered in this paper
(see also Ref. \cite{KK2}), then a condensate
of massive excitation quanta of oscillations of primordial matter which 
behaves as an antigravitating medium can play the role of
the dark energy. Since this condensate is not 
a vacuum, the well-known contradiction with quantum field theory does not arise.

At the end of the paper we note that the voids discovered recently in distant regions 
of space which do not contain ordinary matter \cite{Spot} may be filled with a 
condensate of primordial matter quanta described above that has an antigravitating 
effect on matter produced by this condensate.

\end{document}